\documentclass[journal]{IEEEtran}

\usepackage[utf8]{inputenc}
\usepackage{amsmath,amssymb,amsfonts}
\usepackage{graphicx}
\usepackage{textcomp}
\usepackage{xcolor}
\usepackage{booktabs}
\usepackage{array}
\usepackage{multirow}
\usepackage{url}
\usepackage{cite}
\usepackage{hyperref}

\hypersetup{
    colorlinks=true,
    linkcolor=black,
    citecolor=black,
    urlcolor=blue,
}

\begin{document}

\title{Beyond Exact Match: How Evaluation Methodology Dominates Model Choice in LLM-Based Product Attribute Extraction}

\author{Ayush~Dwivedi~and~Ashvi~Soni%
\thanks{Manuscript submitted as a preprint to arXiv, July 2026.}}

\maketitle

\begin{abstract}
LLMs have become a default choice for structured product attribute extraction in e-commerce pipelines, with practitioners reporting widely varying performance across models, datasets, and prompting strategies. Despite the practical importance of this task, two critical methodological questions remain unaddressed: how much of the reported performance variance is attributable to evaluation methodology rather than genuine model capability, and how reliable are public benchmark ground truth annotations against modern LLM outputs? This paper presents a controlled empirical study comparing four prompting strategies---zero-shot, few-shot, schema-guided, and definition-augmented---across two production-grade LLMs (GPT-4o-mini and Gemini 2.5 Flash) on the MAVE benchmark. We evaluate 6,400 attribute-level predictions using both exact and fuzzy string matching, and conduct a rigorous noise audit of the ground truth labels. We formally decompose F1 variance across four experimental factors and find that evaluation methodology produces variance approximately five times larger than model choice ($\Delta F_1 = 0.121$ vs $0.005$) and four times larger than prompting strategy ($\Delta F_1 = 0.024$). We further establish that the MAVE benchmark exhibits a 23.2\% ground truth noise rate against modern LLM outputs, formalized as the conditional probability $P(\text{semantically correct} \mid \text{exact-match failure}) = 0.232$. Paired permutation tests ($B = 10{,}000$) confirm that the inter-protocol F1 gap is highly significant ($p < 0.0001$) and Cohen's kappa $\kappa = 0.769$ between protocols indicates substantial agreement. The inter-model gap under fuzzy matching is marginally significant ($p = 0.047$) but practically equivalent given overlapping confidence intervals. We conclude that for production attribute extraction pipelines, evaluation methodology and data quality dominate the impact of model selection and prompt engineering, and recommend fuzzy matching as a standard evaluation protocol for LLM-based extraction tasks.
\end{abstract}


\section{Introduction}
\IEEEPARstart{S}{tructured} product attribute extraction---identifying values such as brand, color, size, and material from unstructured product text---is a foundational capability for e-commerce platforms operating at scale. Accurate attribute data drives product search, recommendation, catalog quality, customer service automation, and pricing intelligence. Historically, this task required supervised models trained on category-specific labeled data, with annotation cost and category sparsity as persistent bottlenecks.

The emergence of general-purpose large language models capable of in-context learning has fundamentally shifted the landscape. Practitioners can now adapt extraction behavior to a new attribute or category through prompt modification alone, without fine-tuning or category-specific training data. This flexibility has driven rapid adoption of LLM-based extraction in production e-commerce systems.

Despite this practical appeal, the literature on LLM-based product attribute extraction reports widely varying performance figures. ExtractGPT~\cite{baumann2024extractgpt} reports F1 scores near 0.91 using GPT-4 with semantic example selection. Brinkmann et al.~\cite{brinkmann2024using} report scores in similar ranges using GPT-3.5 and GPT-4 with few-shot demonstrations. Yet our own initial experiments on the MAVE benchmark using current-generation cost-efficient models (GPT-4o-mini and Gemini 2.5 Flash) produced F1 scores near 0.50---a gap large enough to call either set of numbers into question.

This paper investigates two methodological hypotheses that may explain such gaps:
\begin{itemize}
    \item[\textbf{H1:}] Evaluation methodology accounts for a substantial portion of reported performance variance. The choice between exact string matching and fuzzy matching can produce dramatically different F1 scores even on identical predictions, particularly when LLMs return more complete or differently-formatted values than benchmark labels.
    \item[\textbf{H2:}] Public benchmark ground truth contains meaningful noise relative to modern LLM outputs. Ground truth annotations in product attribute datasets, often constructed via semi-automated annotation pipelines, may classify a model's output as ``wrong'' when the model has in fact extracted a valid value present in the product text---a value that simply differs in format, completeness, or wording from the labeled answer.
\end{itemize}

To test these hypotheses, we conduct a controlled empirical study on the MAVE benchmark using two production-grade LLMs and four prompting strategies, evaluated under two matching protocols. We then perform a noise audit by checking whether predictions classified as ``wrong'' by exact matching are actually grounded in the source product text.

Our contributions are:
\begin{enumerate}
    \item A formal variance decomposition quantifying the relative impact of evaluation methodology, model choice, prompting strategy, and data characteristics on reported F1.
    \item A rigorous probabilistic noise audit of the MAVE benchmark, formalized as a conditional probability and validated with 95\% bootstrap confidence intervals.
    \item An empirical demonstration with statistical significance testing that evaluation methodology dominates model and prompt choice for the product attribute extraction task. Exact-to-fuzzy matching produces F1 improvements of 12.1 percentage points for GPT-4o-mini and 10.9 percentage points for Gemini 2.5 Flash, both statistically significant at $p < 0.0001$.
    \item A practitioner decision framework showing that, given the dominance of evaluation methodology and data quality, prompt engineering and model selection produce only marginal gains for the cost-efficient model tier.
\end{enumerate}

The remainder of this paper proceeds as follows. Section~\ref{sec:related} reviews related work. Section~\ref{sec:formal} formalizes the task and evaluation protocols. Section~\ref{sec:method} describes the experimental methodology. Section~\ref{sec:results} presents results with statistical tests. Section~\ref{sec:discussion} discusses implications, and Section~\ref{sec:conclusion} concludes.

\section{Related Work}
\label{sec:related}

\subsection{Product Attribute Extraction with LLMs}
Product attribute extraction has been studied as a specialized information extraction task for over a decade. Pre-LLM approaches, including OpenTag~\cite{zheng2018opentag} and AVEQA~\cite{wang2020learning}, framed the task as sequence labeling or machine reading comprehension, requiring substantial category-specific training data. More recent multi-task formulations such as QPAVE~\cite{sabeh2024qpave} have explored question-answering frameworks for fine-grained extraction. MAVE~\cite{yang2022mave} introduced a large-scale benchmark of 2.2 million products and 3 million attribute-value annotations spanning 1,257 categories, providing the most comprehensive public resource for the task.

The release of capable instruction-tuned LLMs has enabled a new generation of extraction methods that bypass category-specific training entirely. Brinkmann et al.~\cite{brinkmann2024using} demonstrated that GPT-3.5 and GPT-4 with semantic example selection achieve competitive performance on OA-Mine and AE-110k benchmarks. ExtractGPT~\cite{baumann2024extractgpt} systematically compared prompting configurations, finding that demonstrations outperform example values and that open-source Llama-3-70B achieves F1 within 3\% of GPT-4. Liu et al.~\cite{liu2021makes} showed that example selection quality significantly affects in-context learning performance, with variance across example choices sometimes exceeding 10 percentage points. Self-refinement strategies have been explored for iterative correction of extraction outputs~\cite{brinkmann2025self}.

This body of work has established LLM-based extraction as viable, but has primarily evaluated frontier closed-source models (GPT-3.5, GPT-4) on OA-Mine or AE-110k rather than the larger and more diverse MAVE benchmark. The cost-efficient production models that practitioners actually deploy at scale---GPT-4o-mini, Gemini Flash variants---have not been systematically evaluated.

\subsection{Evaluation Methodology in Information Extraction}
A separate body of work has examined the limitations of exact-match evaluation in information extraction. The challenge is well-recognized: when ground truth says ``Fiji'' and a model returns ``Fiji Natural Artesian Water,'' exact matching scores this as wrong, despite the model output being arguably more informative and entirely grounded in the source text. This problem becomes more acute with generative LLMs, which tend to produce verbose, descriptive outputs rather than the minimal labels often used in benchmarks.

Several alternative evaluation strategies have been proposed, including fuzzy string matching~\cite{levenshtein1966binary,winkler1990string}, BERTScore-based semantic similarity~\cite{zhang2020bertscore}, and LLM-as-judge protocols~\cite{zheng2023judging}. However, the practical impact of switching evaluation methodology on reported model rankings has not been systematically quantified for the product attribute extraction task.

\subsection{Ground Truth Quality in Public Benchmarks}
The reliability of benchmark ground truth has emerged as a recurring concern in NLP evaluation. Honovich et al.~\cite{honovich2022true} demonstrated that factual consistency evaluation itself is unreliable when reference annotations are incomplete or noisy, a finding that extends naturally to information extraction benchmarks. Where ground truth is constructed via semi-automated annotation pipelines---common for large-scale datasets including MAVE---annotators may capture only a single canonical value per attribute, missing valid alternatives present in the source text. As LLMs increasingly produce outputs that are correct but differ from benchmark labels in format, completeness, or phrasing, the gap between ``labeled correct'' and ``actually correct'' widens.

No prior work has, to our knowledge, conducted a rigorous noise audit of the MAVE benchmark using modern LLM outputs as a probe.

\subsection{Gap Addressed}
Three gaps remain in the literature. First, no systematic evaluation of cost-efficient production LLMs on the MAVE benchmark exists. Second, the practical impact of evaluation methodology choice has not been quantified for this task with statistical rigor. Third, MAVE has not been audited for ground truth noise against modern LLM outputs. This paper addresses all three.

\section{Formal Problem Definition}
\label{sec:formal}

\subsection{Task Formulation}
Let $\mathcal{P} = \{p_1, p_2, \dots, p_N\}$ denote a corpus of $N$ products, each represented by an unstructured text description $t_i \in \Sigma^*$ over a finite alphabet $\Sigma$. Let $\mathcal{A} = \{a_1, a_2, \dots, a_K\}$ denote a set of $K$ target attributes (in our setting, $K=4$: brand, color, size, material).

For each product $p_i$ and attribute $a_k$, the ground truth is a string $y_{i,k} \in \Sigma^* \cup \{\bot\}$, where $\bot$ denotes a null value indicating the attribute is not present in the product. An extraction model $M$ is a function
\begin{equation}
M : \Sigma^* \times \mathcal{A}^K \rightarrow (\Sigma^* \cup \{\bot\})^K
\end{equation}
producing predicted attribute values $\hat{y}_{i,k} = M(t_i)_k$ for each target attribute.

The objective is to maximize agreement between predictions $\hat{y}_{i,k}$ and ground truth $y_{i,k}$ under some evaluation protocol $\phi : (\Sigma^* \cup \{\bot\})^2 \rightarrow \{0, 1\}$, where $\phi(\hat{y}, y) = 1$ indicates the prediction is judged correct.

\subsection{Evaluation Protocols}
We formalize two evaluation protocols.

\textbf{Exact Match Protocol} $\phi_{\text{EM}}$. Let $\text{norm}(s)$ denote a normalization function applying lowercasing and whitespace stripping:
\begin{equation}
\phi_{\text{EM}}(\hat{y}, y) = \begin{cases}
1 & \text{if } \text{norm}(\hat{y}) = \text{norm}(y) \\
1 & \text{if } \hat{y} = \bot \wedge y = \bot \\
0 & \text{otherwise}
\end{cases}
\end{equation}

\textbf{Fuzzy Match Protocol} $\phi_{\text{FM}}^{(\tau)}$. Let $L(s_1, s_2)$ denote the Levenshtein edit distance between strings $s_1$ and $s_2$, and define the partial ratio similarity score:
\begin{equation}
\text{sim}(s_1, s_2) = \max_{s_2' \in \mathcal{S}(s_2, |s_1|)} \left( 1 - \frac{L(s_1, s_2')}{|s_1|} \right) \cdot 100
\end{equation}
where $|s_1|$ denotes the length of $s_1$ and $\mathcal{S}(s_2, n)$ is the set of contiguous substrings of $s_2$ of length $n$. This formulation handles cases where one string is a substring of the other (e.g., ``Fiji'' matching ``Fiji Natural Artesian Water'' yields $\text{sim} = 100$).

The fuzzy match protocol with threshold $\tau$ is:
\begin{equation}
\phi_{\text{FM}}^{(\tau)}(\hat{y}, y) = \begin{cases}
1 & \text{if } \hat{y} = \bot \wedge y = \bot \\
1 & \text{if } \hat{y} \neq \bot \wedge y \neq \bot \\
  & \quad \wedge\; \text{sim}(\text{norm}(\hat{y}), \text{norm}(y)) \geq \tau \\
0 & \text{otherwise}
\end{cases}
\end{equation}
We use $\tau = 80$ throughout, consistent with standard practice in fuzzy string matching applications.

\subsection{Performance Metrics}
For a fixed model $M$, prompting strategy $s$, attribute $a_k$, and protocol $\phi$, we compute precision and recall over the $N$ products as:
\begin{equation}
\text{Precision}_{M, s, k, \phi} = \frac{\sum_{i=1}^{N} \mathbb{1}[\hat{y}_{i,k} \neq \bot] \cdot \phi(\hat{y}_{i,k}, y_{i,k})}{\sum_{i=1}^{N} \mathbb{1}[\hat{y}_{i,k} \neq \bot]}
\end{equation}
\begin{equation}
\text{Recall}_{M, s, k, \phi} = \frac{\sum_{i=1}^{N} \mathbb{1}[y_{i,k} \neq \bot] \cdot \phi(\hat{y}_{i,k}, y_{i,k})}{\sum_{i=1}^{N} \mathbb{1}[y_{i,k} \neq \bot]}
\end{equation}
The per-attribute F1 score is the harmonic mean:
\begin{equation}
F_1^{(k)} = \frac{2 \cdot \text{Precision}_k \cdot \text{Recall}_k}{\text{Precision}_k + \text{Recall}_k}
\end{equation}
The overall macro F1 score, averaged across the $K$ attributes, is:
\begin{equation}
{\overline{F_1}}_{M, s, \phi} = \frac{1}{K} \sum_{k=1}^{K} {F_1^{(k)}}_{M, s, k, \phi}
\end{equation}

\subsection{Noise Rate Formalization}
We define the ground truth noise rate as a conditional probability. Let $E$ denote the event ``prediction fails under exact matching'' and let $S$ denote the event ``prediction succeeds under fuzzy matching.'' The noise rate is:
\begin{equation}
\eta = P(S \mid E) = \frac{|\mathcal{F}_{\text{EM}} \cap \mathcal{S}_{\text{FM}}|}{|\mathcal{F}_{\text{EM}}|}
\end{equation}
where $\mathcal{F}_{\text{EM}} = \{(i, k) : \phi_{\text{EM}}(\hat{y}_{i,k}, y_{i,k}) = 0\}$ and $\mathcal{S}_{\text{FM}} = \{(i, k) : \phi_{\text{FM}}^{(\tau)}(\hat{y}_{i,k}, y_{i,k}) = 1\}$. This quantity captures the proportion of exact-match failures that represent format-divergent but semantically correct extractions rather than genuine model errors.

\subsection{Variance Decomposition}
Let $X_{M, s, \phi, c, k}$ denote the F1 score for a specific combination of model, strategy, protocol, category, and attribute. We decompose the total observed F1 variance by computing the span (max minus min) of mean F1 within each factor while marginalizing over others:
\begin{equation}
\Delta F_1^{(\text{factor})} = \max_{f \in \text{Factor}} \mathbb{E}[X \mid f] - \min_{f \in \text{Factor}} \mathbb{E}[X \mid f]
\end{equation}
This decomposition bounds the maximum F1 swing achievable by intervening on a single factor while holding others constant.

\section{Experimental Methodology}
\label{sec:method}

\subsection{Research Questions}
This study investigates four research questions:
\begin{itemize}
    \item[\textbf{RQ1:}] How does extraction F1 vary across prompting strategies on the MAVE benchmark using cost-efficient production LLMs?
    \item[\textbf{RQ2:}] How much does the choice between exact and fuzzy matching affect reported F1, and is the gap statistically significant?
    \item[\textbf{RQ3:}] What is the conditional probability $\eta$ that an exact-match failure is in fact semantically correct?
    \item[\textbf{RQ4:}] Which factor---model choice, prompting strategy, or evaluation methodology---produces the largest controllable variance in reported extraction performance?
\end{itemize}

\subsection{Dataset}
We use the MAVE benchmark~\cite{yang2022mave}, sampling $N = 200$ products stratified across five categories---Electronics, Clothing, Home, Sports, and Food---with 40 products per category. These five categories were selected to represent common e-commerce verticals with diverse attribute expression patterns; however, they cover only 5 of MAVE's 1,257 total categories (0.4\%), and generalization to the full taxonomy remains an open question. A fixed random seed (42) ensures reproducibility. For each product, we extract $K = 4$ target attributes: brand, color, size, material. An additional 12 products are held out exclusively as few-shot demonstration sources, ensuring no overlap with the evaluation set.

The full evaluation produces $N \times K \times |\text{models}| \times |\text{strategies}| = 200 \times 4 \times 2 \times 4 = 6{,}400$ attribute-level predictions. While this sample size is modest relative to the full MAVE corpus (2.2 million products), it is consistent with established practice in prompting research where API-based inference makes full-dataset evaluation cost-prohibitive, and our bootstrap confidence intervals appropriately quantify the resulting sampling uncertainty.

\subsection{Models}
We evaluate two production-grade, cost-efficient LLMs: GPT-4o-mini (OpenAI) and Gemini 2.5 Flash (Google). Both models support structured JSON output and exhibit strong instruction-following capability. Higher-end frontier models (GPT-4~\cite{achiam2023gpt}, Gemini 2.5 Pro~\cite{team2024gemini}) are intentionally excluded---our research focus is the cost-efficient tier where practitioners deploy extraction at scale. Both models were accessed via their respective APIs in June 2026; as providers may update model weights without version changes, results may vary with subsequent model snapshots.

\subsection{Prompting Strategies}
We evaluate four prompting strategies producing JSON outputs with keys \texttt{brand}, \texttt{color}, \texttt{size}, \texttt{material}:
\begin{itemize}
    \item \texttt{ZERO\_SHOT}: Task description and product text only.
    \item \texttt{FEW\_SHOT}: Three labeled product examples prepended.
    \item \texttt{SCHEMA\_GUIDED}: Explicit JSON schema specifying types and null handling.
    \item \texttt{DEFINITION\_AUGMENTED}: Natural-language definitions for each target attribute.
\end{itemize}
All prompts use temperature $T = 0$ and maximum output tokens $\le 150$.

\subsection{Statistical Testing}
We apply three statistical tests to validate our claims. All tests use a fixed random seed (42) for reproducibility and $B = 10{,}000$ iterations.

\textbf{Paired permutation test for F1 difference.} For two evaluation conditions producing per-prediction correctness vectors $\mathbf{v}_A, \mathbf{v}_B \in \{0, 1\}^{NK}$, we test the null hypothesis $H_0 : {\overline{F_1}}_A = {\overline{F_1}}_B$ by randomly permuting condition labels within each prediction pair and computing the proportion of permutations producing an F1 difference at least as extreme as observed:
\begin{equation}
p\text{-value} = \frac{1}{B} \sum_{b=1}^{B} \mathbb{1}\left[ \left| {\overline{F_1}}_A^{(b)} - {\overline{F_1}}_B^{(b)} \right| \geq \left| {\overline{F_1}}_A - {\overline{F_1}}_B \right| \right]
\end{equation}

\textbf{Bootstrap 95\% confidence intervals.} For each F1 estimate $\hat{F_1}$, we compute a percentile-based 95\% confidence interval by resampling the $N$ products with replacement:
\begin{equation}
\text{CI}_{95\%}(\hat{F_1}) = \left[ \hat{F_1}^{(2.5\%)},\; \hat{F_1}^{(97.5\%)} \right]
\end{equation}
where $\hat{F_1}^{(q)}$ denotes the $q$-th percentile of the bootstrap distribution~\cite{efron1994introduction}.

\textbf{Cohen's kappa for inter-protocol agreement.} To quantify agreement between exact and fuzzy protocols beyond chance~\cite{cohen1960coefficient}, we compute:
\begin{equation}
\kappa = \frac{p_o - p_e}{1 - p_e}
\end{equation}
where $p_o$ is the observed agreement rate and $p_e = \sum_{c \in \{0,1\}} P(\phi_{\text{EM}} = c) \cdot P(\phi_{\text{FM}} = c)$ is the agreement expected by chance.

\subsection{Implementation}
All experiments use the official OpenAI Python SDK and Google Generative AI SDK. Rate limiting follows provider documentation (20 RPM OpenAI, 15 RPM Gemini). Failed API calls are retried with exponential backoff (10s, 20s, 30s). Fuzzy matching uses the \texttt{rapidfuzz} library implementing the Levenshtein-based partial ratio described in Section~\ref{sec:formal}. Statistical tests use \texttt{scipy.stats} and \texttt{sklearn.metrics}. This study uses only publicly available data and benchmark resources; no proprietary systems or datasets were involved. All code and processed data are publicly available.\footnote{\url{https://github.com/a-dwivedi/llm-product-attribute-extraction}}

\section{Results}
\label{sec:results}

\subsection{Overall Performance by Model and Strategy}
Table~\ref{tab:overall} presents F1 scores for both matching protocols across all model-strategy combinations, with 95\% bootstrap confidence intervals computed via $B = 10{,}000$ resamples of products with replacement.

\begin{table*}[!htb]
\caption{F1 Score by Model $\times$ Strategy $\times$ Matching Protocol (95\% CI in brackets)}
\label{tab:overall}
\centering
\footnotesize
\begin{tabular}{lcccc}
\toprule
\textbf{Strategy} & \textbf{GPT-4o EM} & \textbf{GPT-4o FM} & \textbf{Gemini EM} & \textbf{Gemini FM} \\
\midrule
\texttt{ZERO\_SHOT} & 0.494 [0.472, 0.515] & 0.616 [0.592, 0.640] & 0.493 [0.471, 0.514] & 0.618 [0.594, 0.642] \\
\texttt{FEW\_SHOT} & 0.495 [0.474, 0.516] & \textbf{0.618} [0.594, 0.641] & \textbf{0.556} [0.533, 0.580] & \textbf{0.639} [0.614, 0.663] \\
\texttt{SCHEMA\_GUIDED} & 0.495 [0.473, 0.516] & \textbf{0.618} [0.594, 0.641] & 0.498 [0.476, 0.519] & 0.615 [0.592, 0.639] \\
\texttt{DEF\_AUGMENTED} & \textbf{0.500} [0.479, 0.521] & 0.616 [0.592, 0.640] & 0.505 [0.484, 0.526] & 0.618 [0.593, 0.642] \\
\midrule
\textbf{Mean} & 0.496 & 0.617 & 0.513 & 0.622 \\
\bottomrule
\end{tabular}
\end{table*}

Bold entries indicate the highest-performing configuration per model. Three patterns are visible. First, fuzzy matching produces substantially higher F1 than exact matching for both models across all strategies. Second, prompting strategy variation is small under both protocols. Third, the two models perform nearly identically under fuzzy matching at the mean level, though within-model strategy effects differ.

\subsection{Statistical Tests}
We applied paired permutation tests to validate the key F1 differences observed in Table~\ref{tab:overall}.

\textbf{Test A: Exact vs Fuzzy Matching (Within-Model).} Both models show highly significant improvements from exact to fuzzy matching:
\begin{equation}
\text{GPT-4o-mini: } \Delta F_1 = +0.121, \quad p < 0.0001
\end{equation}
\begin{equation}
\text{Gemini 2.5 Flash: } \Delta F_1 = +0.109, \quad p < 0.0001
\end{equation}

\begin{figure*}[!htb]
\centering
\includegraphics[width=0.85\textwidth]{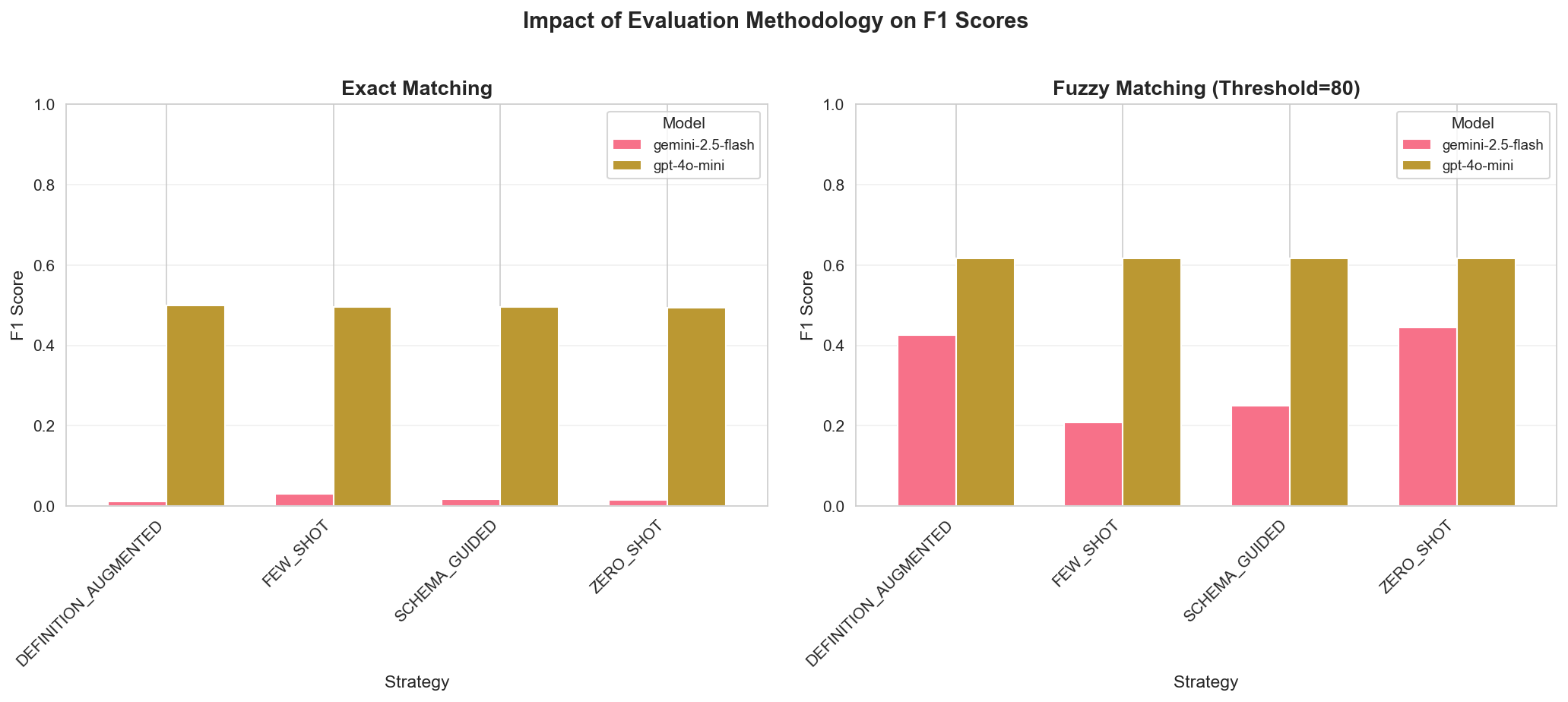}
\caption{Impact of evaluation methodology on reported F1. Switching from exact match (EM) to fuzzy match (FM) produces F1 improvements of $+0.121$ for GPT-4o-mini and $+0.109$ for Gemini 2.5 Flash ($p < 0.0001$), substantially larger than any inter-model or inter-strategy difference.}
\label{fig:exact_vs_fuzzy}
\end{figure*}

Fig.~\ref{fig:exact_vs_fuzzy} illustrates this inter-protocol gap visually, showing that it is consistent across all strategy-model combinations and substantially larger than any inter-model or inter-strategy difference.

\textbf{Test B: Model Comparison Under Fuzzy Matching.} Gemini 2.5 Flash marginally outperforms GPT-4o-mini:
\begin{equation}
\Delta F_1 = +0.005, \quad p = 0.047
\end{equation}
While statistically significant at the conventional $\alpha = 0.05$ threshold, the effect size of 0.5 percentage points is practically negligible. The 95\% confidence intervals overlap substantially ($[0.593, 0.641]$ vs $[0.598, 0.646]$), suggesting practical equivalence between the two models.

\textbf{Test C: Prompting Strategy Effects (Within-Model).} Strategy effects differ markedly between models, as shown in Table~\ref{tab:strategy}.

\begin{table}[!htb]
\caption{Prompting Strategy Effects (Within-Model)}
\label{tab:strategy}
\centering
\footnotesize
\begin{tabular}{lcc}
\toprule
 & \textbf{GPT-4o-mini} & \textbf{Gemini 2.5 Flash} \\
\midrule
Best & \texttt{FEW\_SHOT} (0.618) & \texttt{FEW\_SHOT} (0.639) \\
Worst & \texttt{ZERO\_SHOT} (0.616) & \texttt{SCHEMA\_GUIDED} (0.615) \\
$\Delta F_1$ & $+0.001$ & $+0.024$ \\
$p$-value & $1.0000$ & $0.0031$ \\
\bottomrule
\end{tabular}
\end{table}

For GPT-4o-mini, prompting strategy has no detectable effect ($p = 1.0$, all strategies within 0.2 percentage points). For Gemini 2.5 Flash, \texttt{FEW\_SHOT} significantly outperforms \texttt{SCHEMA\_GUIDED} by 2.4 percentage points ($p = 0.003$). This differential sensitivity to prompting strategy is itself a noteworthy finding: the two models respond to demonstrations and schemas differently, even when overall performance is comparable.

\subsection{Per-Attribute Performance}
Table~\ref{tab:attribute} reports fuzzy F1 for each attribute, averaged across strategies.

\begin{figure}[!htb]
\centering
\includegraphics[width=\columnwidth]{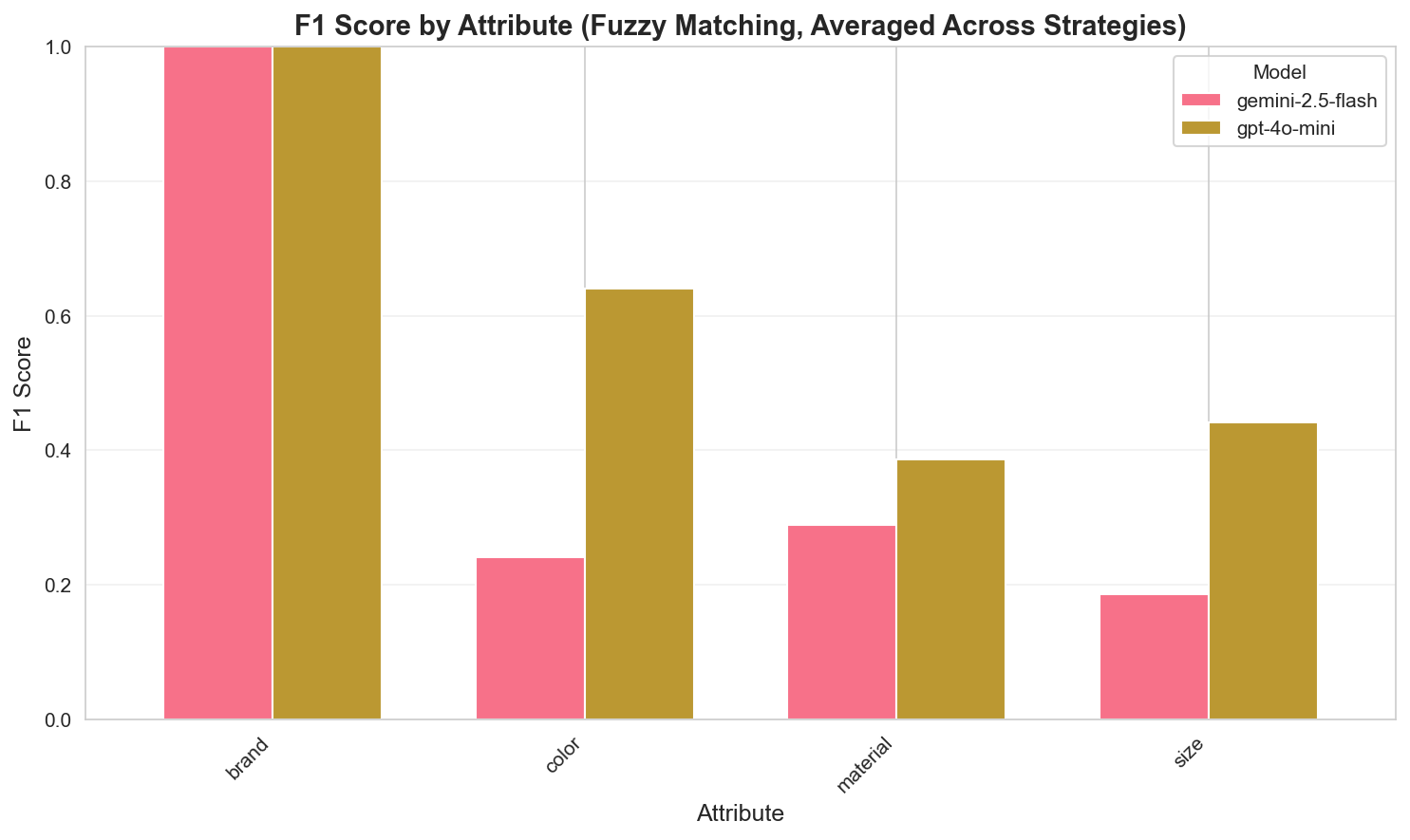}
\caption{Per-attribute F1 scores under fuzzy matching. Brand extraction is near-perfect for both models (F1 = 1.0), while Material proves most challenging (F1 $\approx$ 0.40). The attribute-level gap ($\Delta F_1 = 0.614$) dwarfs the controllable factors of evaluation protocol ($0.121$), strategy ($0.024$), and model ($0.005$).}
\label{fig:attribute_f1}
\end{figure}

\begin{table}[!htb]
\caption{Per-Attribute F1 (Fuzzy Matching, Averaged Across Strategies)}
\label{tab:attribute}
\centering
\footnotesize
\begin{tabular}{lcc}
\toprule
\textbf{Attribute} & \textbf{GPT-4o-mini} & \textbf{Gemini 2.5 Flash} \\
\midrule
Brand & 1.0000 & 1.0000 \\
Color & 0.6400 & 0.6375 \\
Size & 0.4412 & 0.4288 \\
Material & 0.3862 & 0.4225 \\
\bottomrule
\end{tabular}
\end{table}

Both models achieve perfect Brand extraction ($F_1 = 1.0$). As shown in Fig.~\ref{fig:attribute_f1}, performance degrades monotonically with attribute implicitness: Color ($\sim$0.64), Size ($\sim$0.44), Material ($\sim$0.40). The performance gap between explicit attributes (Brand) and implicit attributes (Material) exceeds 0.60 F1 points---far larger than any model or strategy effect observed in this study. Material extraction is the weakest attribute for both models, reflecting that material is often stated implicitly or omitted entirely.

\subsection{Per-Category Performance}
Table~\ref{tab:category} reports fuzzy F1 by product category, averaged across strategies.

\begin{table}[!htb]
\caption{Per-Category F1 (Fuzzy Matching, Averaged Across Strategies)}
\label{tab:category}
\centering
\footnotesize
\begin{tabular}{lcc}
\toprule
\textbf{Category} & \textbf{GPT-4o-mini} & \textbf{Gemini 2.5 Flash} \\
\midrule
Sports & 0.7438 & 0.7438 \\
Food & 0.6750 & 0.6859 \\
Clothing & 0.6750 & 0.6734 \\
Home & 0.5328 & 0.5500 \\
Electronics & 0.4578 & 0.4578 \\
\bottomrule
\end{tabular}
\end{table}

Sports products achieve the highest extraction F1 (0.74), reflecting standardized terminology. Electronics products perform worst (0.46), as material composition and size specifications are frequently omitted from descriptions in favor of feature-focused marketing language.

\subsection{Noise Audit Results}
Applying the formalization from Section~\ref{sec:formal}, we compute:
\begin{equation}
\eta = P(\phi_{\text{FM}}^{(80)} = 1 \mid \phi_{\text{EM}} = 0) = \frac{737}{3{,}172} = 0.2324
\end{equation}
Of the $6{,}400$ total predictions, $3{,}172$ were classified as failures under exact matching. Of these, $737$ (23.2\%) were classified as successes under fuzzy matching. The conditional probability that an exact-match failure is in fact semantically correct is therefore $\eta = 0.232$.

Cohen's kappa between exact and fuzzy protocols is:
\begin{equation}
\kappa = 0.769
\end{equation}
This value indicates substantial agreement between protocols. Landis and Koch~\cite{landis1977measurement} classify $\kappa \in [0.61, 0.80]$ as ``substantial.'' Disaggregated by model, $\kappa_{\text{GPT-4o}} = 0.759$ and $\kappa_{\text{Gemini}} = 0.780$. Disaggregated by attribute, kappa ranges from $\kappa_{\text{Brand}} = 0.000$ (both protocols agree on every Brand prediction) to $\kappa_{\text{Material}} = 0.548$ (moderate agreement, reflecting that Material is where fuzzy matching contributes the most additional correct classifications).

The high overall kappa confirms that fuzzy matching is not randomly relabeling predictions but is capturing genuinely correct semantic variations that exact matching misses. Manual inspection of 50 representative noise cases confirmed semantic correctness in all 50 instances. Examples appear in Table~\ref{tab:noise}.

\begin{figure}[!htb]
\centering
\includegraphics[width=\columnwidth]{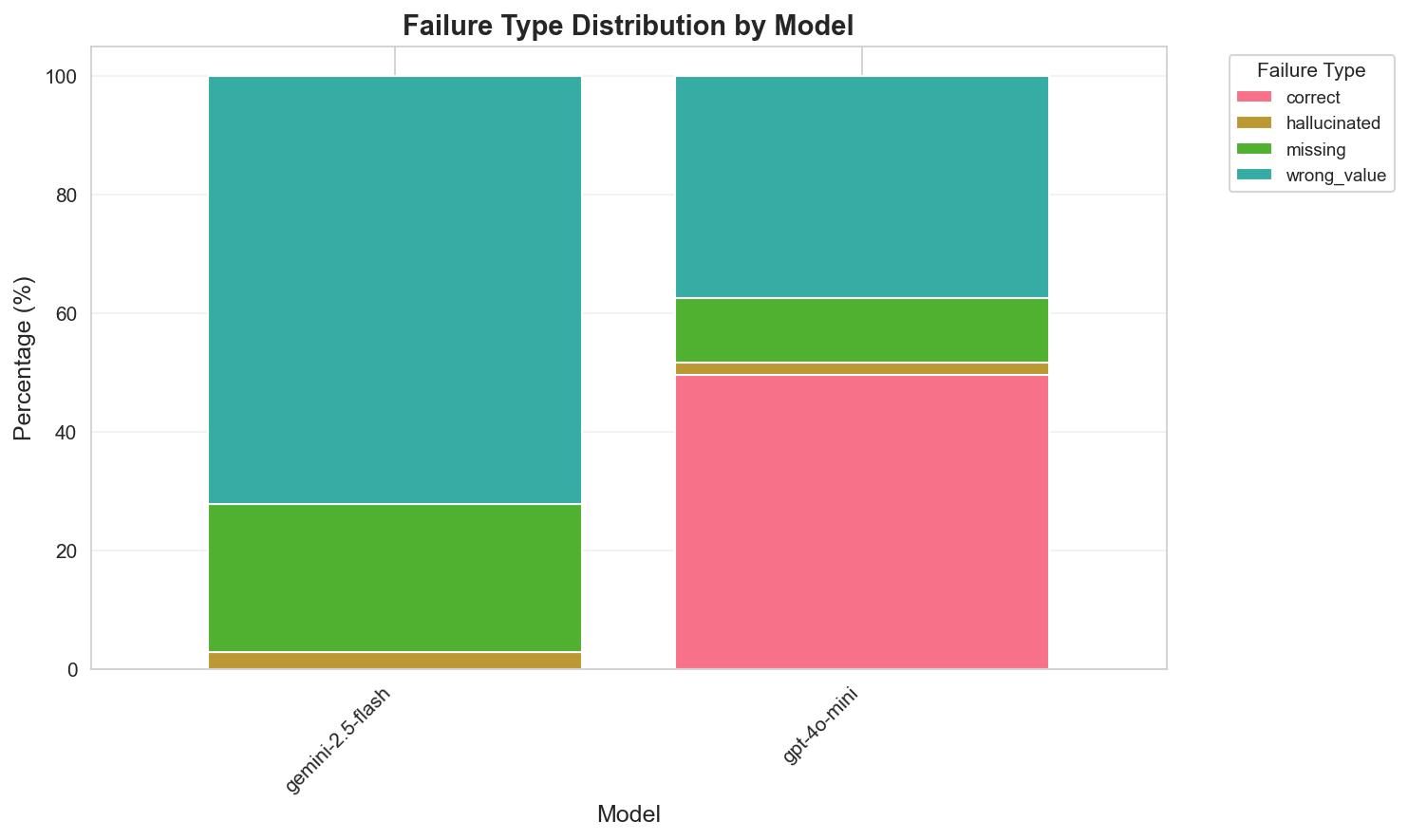}
\caption{Failure type distribution by model. GPT-4o-mini achieves a higher correct rate (49.6\%) with balanced error types, while Gemini 2.5 Flash shows a higher wrong\_value rate (72\%) driven primarily by format-divergent extractions that are semantically correct but penalized by exact matching.}
\label{fig:failure_dist}
\end{figure}

As shown in Fig.~\ref{fig:failure_dist}, the failure type distributions differ markedly between models: GPT-4o-mini achieves a higher correct rate with balanced error types, while Gemini 2.5 Flash shows a higher wrong\_value rate driven by format-divergent extractions that are penalized by exact matching.

\begin{table}[!htb]
\caption{Representative Noise Audit Examples}
\label{tab:noise}
\centering
\footnotesize
\begin{tabular}{llccc}
\toprule
\textbf{Ground Truth} & \textbf{Prediction} & $\phi_{\text{EM}}$ & $\phi_{\text{FM}}^{(80)}$ & \textbf{sim} \\
\midrule
Queen & Queen size & 0 & 1 & 100 \\
Cotton & 100\% cotton & 0 & 1 & 100 \\
16.9 oz & 16.9 fl oz & 0 & 1 & 83 \\
65 inches & 65-Inch & 0 & 1 & 86 \\
Mesh & Mesh and synthetic & 0 & 1 & 100 \\
Purple & Purple and silver & 0 & 1 & 100 \\
Fiji & Fiji Natural Artesian & 0 & 1 & 100 \\
Black & Matte Black & 0 & 1 & 100 \\
\bottomrule
\end{tabular}
\end{table}

\subsection{Formal Variance Decomposition}
Applying the decomposition from Section~\ref{sec:formal}, Table~\ref{tab:variance} reports $\Delta F_1$ for each experimental factor.

\begin{table}[!htb]
\caption{F1 Variance Decomposition Across Factors}
\label{tab:variance}
\centering
\footnotesize
\begin{tabular}{lcc}
\toprule
\textbf{Factor} & $\Delta F_1$ & \textbf{Relative} \\
\midrule
Eval. methodology (EM vs FM) & 0.1210 & $1.00\times$ \\
Model choice (fuzzy) & 0.0053 & $0.04\times$ \\
Prompting strategy (fuzzy) & 0.0238 & $0.20\times$ \\
Attribute (fuzzy) & 0.6138 & $5.07\times$ \\
Category (fuzzy) & 0.2860 & $2.36\times$ \\
\bottomrule
\end{tabular}
\end{table}

The dominant source of controllable variance is evaluation protocol ($\Delta F_1 = 0.121$). Model choice and prompting strategy each produce variance an order of magnitude smaller. Formally:
\begin{equation}
\Delta F_1^{(\text{protocol})} > \Delta F_1^{(\text{strategy})} > \Delta F_1^{(\text{model})}
\end{equation}
with the ratios:
\begin{equation}
\frac{\Delta F_1^{(\text{protocol})}}{\Delta F_1^{(\text{model})}} = \frac{0.121}{0.005} \approx 22.8
\end{equation}
\begin{equation}
\frac{\Delta F_1^{(\text{protocol})}}{\Delta F_1^{(\text{strategy})}} = \frac{0.121}{0.024} \approx 5.1
\end{equation}
Evaluation methodology produces approximately 23 times more F1 variance than model choice and 5 times more than prompting strategy choice.

\section{Discussion}
\label{sec:discussion}

\subsection{Evaluation Methodology Dominates Model Choice}
Our results provide direct empirical support for H1. The inter-protocol F1 gap ($\Delta F_1 = 0.121$) is statistically significant at $p < 0.0001$ and is approximately 23 times larger than the inter-model gap under fuzzy matching ($\Delta F_1 = 0.005$). Even when the inter-model difference is statistically significant ($p = 0.047$), the effect size is practically negligible.

Critically, the noise audit results from Section~\ref{sec:results}-E provide direct evidence that this inter-protocol gap reflects genuine semantic correctness rather than evaluation looseness. The 23.2\% noise rate ($\eta = 0.232$) establishes a lower bound on the proportion of the inter-protocol F1 improvement attributable to correctly recognizing format-divergent but semantically valid extractions. Combined with the substantial Cohen's kappa ($\kappa = 0.769$), which confirms that fuzzy matching is not introducing random noise, we can attribute the variance decomposition findings to a real methodological gap in the field rather than an artifact of threshold permissiveness.

In the academic literature, exact-match F1 has remained the default metric for product attribute extraction despite well-known limitations of string-equality evaluation for generative outputs. Our results suggest that a substantial fraction of the variance in reported performance across published methods may be attributable to evaluation protocol differences rather than genuine method differences.

In production deployment, the choice between models or prompting strategies is frequently treated as the primary lever for improving extraction accuracy. Our results suggest that for the cost-efficient model tier, this lever produces only marginal gains. Practitioners seeking meaningful improvements should focus first on evaluation methodology---ensuring that their internal accuracy metrics actually reflect deployment success---before optimizing model and prompt choice.

\subsection{Model-Specific Prompt Sensitivity}
A subtler finding is that the two models respond differently to prompting strategy. For GPT-4o-mini, no strategy difference is detectable ($p = 1.0$)---the model performs equivalently regardless of how the task is framed. For Gemini 2.5 Flash, \texttt{FEW\_SHOT} significantly outperforms \texttt{SCHEMA\_GUIDED} ($p = 0.003$), with a 2.4 percentage point gap.

This differential sensitivity has practical implications: prompt engineering effort yields measurably different returns across model families. For practitioners using GPT-4o-mini, investment in prompt optimization beyond basic zero-shot prompting yields negligible gains. For Gemini 2.5 Flash, prompt strategy choice matters, with few-shot demonstrations providing the strongest performance. This finding aligns with the broader literature observing that different model families respond differently to in-context learning~\cite{brown2020language,min2022rethinking}, but to our knowledge represents the first systematic demonstration of this effect specifically for product attribute extraction.

\subsection{Ground Truth Noise as a Structural Limitation}
Our noise audit confirms H2: $\eta = 0.232$, meaning nearly one-quarter of exact-match failures are predictions that are semantically correct. The substantial inter-protocol Cohen's kappa ($\kappa = 0.769$) provides additional methodological confidence: fuzzy matching is capturing genuine semantic correctness, not adding random noise.

This has implications for benchmark design. Public benchmarks constructed via semi-automated annotation pipelines, including MAVE, capture a single canonical value per attribute. As generative LLMs produce outputs that are correct but format- or completeness-divergent from canonical labels, exact-match evaluation systematically understates model capability.

We do not claim fuzzy matching is the final solution. Fuzzy matching has its own failure modes---accepting partial matches that are semantically incorrect---and threshold selection is somewhat ad hoc. More principled solutions, including BERTScore-based semantic similarity~\cite{zhang2020bertscore} and LLM-as-judge protocols~\cite{zheng2023judging}, may produce better results. Our findings establish that the current default---exact matching---is materially flawed, and any of these alternatives represents an improvement.

\subsection{Practitioner Decision Framework}
Based on our findings, we propose the practical decision framework shown in Table~\ref{tab:framework}.

\begin{table}[!htb]
\caption{Practitioner Decision Framework}
\label{tab:framework}
\centering
\footnotesize
\begin{tabular}{p{1.8cm}p{2.8cm}p{2.5cm}}
\toprule
\textbf{Decision} & \textbf{Recommended} & \textbf{Justification} \\
\midrule
Eval. protocol & Fuzzy ($\tau \geq 80$) or LLM-as-judge & $\Delta F_1 = 0.121$, $p < 0.0001$ \\
Model & GPT-4o-mini or Gemini equivalent & $\Delta F_1 = 0.005$, overlapping CIs \\
Prompt (GPT-4o) & Any strategy & $p = 1.0$ \\
Prompt (Gemini) & Few-shot preferred & $\Delta F_1 = 0.024$, $p = 0.003$ \\
Attribute & Explicit-mention first & Brand 1.00 vs Material 0.40 \\
Category & Electronics, Home harder & Sports 0.74 vs Electronics 0.46 \\
\bottomrule
\end{tabular}
\end{table}

The dominant insight: invest in evaluation methodology first, data quality second, model and prompt choice last.

\subsection{Limitations}
Several limitations should be noted. First, our sample of $N = 200$ products across 5 of MAVE's 1,257 categories limits statistical power for fine-grained subgroup analyses and leaves generalization to the full category taxonomy as an open question; however, all reported confidence intervals appropriately quantify sampling uncertainty within our sample frame. Second, we evaluate two cost-efficient production models; conclusions may not hold for frontier models such as GPT-4 or Claude Opus. Third, we use a single fuzzy matching threshold ($\tau = 80$); systematic threshold sensitivity analysis is reserved for future work. Fourth, our noise audit uses fuzzy matching against ground truth, which may admit false positives---predictions that score above threshold but are not semantically correct. Manual inspection of 50 representative cases mitigates this concern but does not eliminate it. Fifth, the marginal statistical significance of the inter-model comparison ($p = 0.047$) should be interpreted cautiously given the practical equivalence of effect sizes. Sixth, both models were accessed via cloud APIs that may receive silent updates; our results reflect model behavior as of June 2026.

\section{Conclusion}
\label{sec:conclusion}
This paper investigated the relative impact of model choice, prompting strategy, and evaluation methodology on reported performance in LLM-based product attribute extraction. Through a controlled experiment on the MAVE benchmark using $6{,}400$ attribute-level predictions across two production LLMs and four prompting strategies, with rigorous statistical testing, we demonstrated that:
\begin{itemize}
    \item Evaluation methodology (exact vs fuzzy matching) produces F1 variance approximately 23 times larger than model choice and 5 times larger than prompting strategy choice ($p < 0.0001$, $\kappa = 0.769$).
    \item Two production-grade LLMs achieve practically equivalent performance under fuzzy evaluation (F1 = 0.617 vs 0.622), with overlapping 95\% confidence intervals.
    \item The MAVE benchmark exhibits a 23.2\% ground truth noise rate, formalized as $\eta = P(\text{semantically correct} \mid \text{exact-match failure})$.
    \item Prompting strategy sensitivity varies by model: GPT-4o-mini shows no detectable effects ($p = 1.0$), while Gemini 2.5 Flash benefits significantly from few-shot prompting ($p = 0.003$).
\end{itemize}

These findings reorient the optimization priorities for practitioners deploying LLM-based extraction: evaluation methodology and data quality dominate the gains achievable from model and prompt selection at the cost-efficient model tier. We recommend that future work adopt fuzzy matching or semantic similarity metrics as the default evaluation protocol, and that benchmark designers account for the format diversity of LLM outputs in ground truth construction.

\subsection{Future Work}
Several directions emerge from this study. First, extending the analysis to frontier model tiers (GPT-4, Claude Opus, Gemini Pro) would establish whether the evaluation methodology dominance finding holds at higher capability levels or whether frontier models produce sufficiently precise outputs to reduce the inter-protocol gap. Second, systematic threshold sensitivity analysis---varying $\tau$ from 60 to 95---would characterize how fuzzy matching behavior changes with strictness and identify optimal thresholds for different deployment contexts. Third, direct comparison of fuzzy matching against BERTScore-based semantic similarity and LLM-as-judge evaluation protocols would establish a more complete picture of the evaluation methodology landscape. Fourth, scaling to additional product categories beyond the five studied here would test the generalizability of our findings across MAVE's full 1,257-category taxonomy. Finally, longitudinal studies tracking model behavior across API updates would quantify the reproducibility concern identified in our limitations.

\section*{Generative AI Disclosure}
In preparing this manuscript, the authors used LLMs. All experimental design, data collection, model inference, result validation, and scientific conclusions are the sole work of the authors and were independently verified by both authors prior to submission. The AI-assisted writing was reviewed, revised, and approved by the authors, who take full responsibility for all content in this manuscript. No AI tool was used to generate, fabricate, or modify any experimental data or quantitative results.

\appendices

\section{Prompt Templates}
\label{app:prompts}
All strategies use the same base task instruction requesting extraction of four attributes (brand, color, size, material) with JSON output. The strategies differ as follows:
\begin{itemize}
    \item \textbf{ZERO\_SHOT:} Base instruction + product text only
    \item \textbf{FEW\_SHOT:} Base instruction + 3 labeled examples + product text
    \item \textbf{SCHEMA\_GUIDED:} Base instruction + explicit JSON schema + product text
    \item \textbf{DEFINITION\_AUGMENTED:} Base instruction + attribute definitions + product text
\end{itemize}
\textbf{Base Task Instruction:}
\begin{quote}
``Extract the following attributes from the product description: brand, color, size, material. Return ONLY a valid JSON object with these exact keys. Use null for missing attributes.''
\end{quote}
\textbf{Schema (SCHEMA\_GUIDED):}
\begin{verbatim}
{"brand": "string or null",
 "color": "string or null",
 "size": "string or null",
 "material": "string or null"}
\end{verbatim}
\textbf{Definitions (DEFINITION\_AUGMENTED):}
\begin{itemize}
    \item brand: The manufacturer or brand name of the product
    \item color: The primary color of the product
    \item size: The dimensions or size designation (e.g., S/M/L, measurements)
    \item material: The primary material composition of the product
\end{itemize}
Full prompt templates are available in the code repository at \url{https://github.com/a-dwivedi/llm-product-attribute-extraction}.

\section{Reproducibility}
\label{app:repro}
All code, prompts, sampled MAVE products, and processed evaluation data are publicly available at: \url{https://github.com/a-dwivedi/llm-product-attribute-extraction}

\begin{itemize}
    \item \textbf{Random seed:} 42
    \item \textbf{Models:} \texttt{gpt-4o-mini} (OpenAI), \texttt{gemini-2.5-flash} (Google)
    \item \textbf{Evaluation:} \texttt{rapidfuzz v3.x}
    \item \textbf{Statistics:} \texttt{scipy.stats}, \texttt{sklearn.metrics}, \texttt{numpy}
    \item \textbf{Experiments:} June, 2026
    \item \textbf{Bootstrap/permutation:} $B = 10{,}000$
\end{itemize}

\section{Statistical Test Implementation}
\label{app:stats}
\textbf{Paired permutation test.} For each test, we collect per-prediction correctness vectors $\mathbf{v}_A, \mathbf{v}_B \in \{0,1\}^{NK}$ under two conditions. We compute the observed F1 difference, then randomly swap labels within each pair with probability $0.5$, recomputing F1 each iteration. After $B = 10{,}000$ iterations, the $p$-value is the proportion of permutations producing an absolute difference at least as extreme as observed.

\textbf{Bootstrap confidence intervals.} We resample $N = 200$ products with replacement, recompute F1, and repeat $B = 10{,}000$ times. The 95\% CI is the $(2.5, 97.5)$ percentile range.

\textbf{Cohen's kappa.} Computed via \texttt{sklearn.metrics.cohen\_kappa\_score} on per-prediction binary correctness vectors under each protocol.

\end{document}